\documentclass[preprint,aps,prd,amsmath,amscd,amssymb]{revtex4}
\usepackage{graphicx}
\usepackage{color}
\usepackage{graphicx}
\usepackage{dcolumn}
\usepackage{bm}
\usepackage{amsfonts}
\usepackage{bbm}
\usepackage{subfigure}
\usepackage[latin1]{inputenc}
\usepackage[T1]{fontenc}
\usepackage[english]{babel}
\usepackage{fancyhdr}
\usepackage{color}
\usepackage{hyperref}
\usepackage{epsfig}
\usepackage{latexsym}
\newcommand{\psib}{\ensuremath{\overline{\psi}}}

\def\drawbox#1#2{\hrule height#2pt
        \hbox{\vrule width#2pt height#1pt \kern#1pt
              \vrule width#2pt}
              \hrule height#2pt}

\def\Asym#1#2{\vcenter{\vbox{\drawbox{#1}{#2}
              \kern-#2pt 
              \drawbox{#1}{#2}}}}

\newcommand{\beq}{\begin{eqnarray}}
\newcommand{\eeq}{\end{eqnarray}}

\newcommand{\bmp}{\noindent\begin{minipage}{16cm}}
\newcommand{\emp}{\end{minipage}\vskip 7mm} 

\begin{document}
\title{{\large Minimal Walking on the Lattice}}

\author{Simon Catterall}

\affiliation{Department of Physics, Syracuse University, USA.
Syracuse, NY 13244-1130 }

\author{Francesco Sannino}
\affiliation{CERN Theory Division, CH-1211 Geneva 23, Switzerland}
\affiliation{University of Southern Denmark, Campusvej 55, DK-5230 Odense M, Denmark.}


\begin{abstract}
We provide the first 
evidence of a walking dynamics
for two color lattice Yang-Mills theory
with two Dirac flavors in the 
symmetric representation of the gauge group.
\end{abstract}


\maketitle
\section{Introduction}

We undertake the 
first numerical study
of a four dimensional asymptotically free 
gauge theory with dynamical fermions transforming according to the two index symmetric representation of the gauge group. 
To be more precise we consider the simplest of these theories with interesting dynamics and phenomenological applications, i.e. a two color gauge theory with two Dirac fermions transforming according to the two index symmetric representation. For two colors the two index symmetric coincides with the adjoint representation and the reality of the representation enhances the quantum flavor symmetry group to $SU(4)$. Remarkably this theory coincides with the fermionic sector of the ${\cal N}=4$ supersymmetric theory with two colors.

Recently it has been argued that this theory lies close to a non trivial infrared fixed point \cite{Sannino:2004qp}. In this case the coupling constant will run very slowly from the infrared to the ultraviolet; the coupling is said
to {\it walk}. By analyzing the phase diagram as function of the number of colors and flavors in $SU(N)$ gauge theories with Dirac fermions in a given arbitrary representation of the gauge group we have shown that this theory is  {\it minimal} in the sense that it is the theory with the smallest number of
flavors  (above one) which exhibits such walking dynamics
\cite{Dietrich:2006cm}.

The walking dynamics is expected to be dramatically different than in the QCD case. This is so since the presence of a nearby I.R. fixed point should generate
an anomalously small mass scale and ensure that
long distance quantities are insensitive to the
short distance coupling. The physics of the fixed point theory per se is very interesting and when coupled to non-conformal theories (such as the standard model) in the way described recently by Georgi \cite{Georgi:2007ek} it leads to interesting experimental signatures. This is so since the presence of a conformal symmetry signals itself experimentally in a way that {\it formally} resembles the production of a non-integer number of massless invisible particles. The non-integer number is nothing but the scale dimension of the operator, coming from the underlying conformal theory, coupled weakly to the standard model operators.  However, as also stressed by Georgi, very little is known about  conformal or near-conformal theories in four dimensions because of the complicated nonperturbative dynamics. {}Our work should be considered as a first step in this direction. 

In addition,
the emergence of a walking property for
just two Dirac flavors renders this theory an ideal candidate to break the electroweak theory dynamically \cite{Sannino:2004qp}. This extension of the standard model passes the stringent electroweak precision constraints \cite{Dietrich:2005jn}. Moreover, it allows for  
a successful unification 
of the standard model couplings \cite{Gudnason:2006mk} and 
allows us to construct different types of dark matter candidates \cite{Kainulainen:2006wq,Kouvaris:2007iq,Gudnason:2006yj}. 

The lattice results presented in this work support
the theoretical expectations, i.e. that the present theory walks.
To arrive to this conclusion we 
analyze various physical quantities on the lattice and then compare them to the ones for a two color gauge theory with two Dirac flavors in the fundamental representation of the gauge group. The latter theory, as it is clear from the phase diagram reviewed later, is very far away from an infrared fixed point. 
Although our simulations employ dynamical quarks
the lattices we use are small so these results should only be taken as
indicative of qualitative agreement with the theoretical expectations
and encourage one to embark on
a more serious study on larger lattices.

In the next section we review the expected phase diagram 
as function of flavors and colors for non supersymmetric asymptotically free $SU(N)$ gauge theories with $N_f$ Dirac fermions in a given representation of the gauge group \cite{Dietrich:2006cm}. Here it is shown that the model
we study in this paper is indeed the theory with the lowest number of Dirac flavors (above one) able to feature walking. This feature makes this theory also an ideal candidate when used for breaking the electroweak theory dynamically. We review the salient features and the notation here.

We then describe the details of our lattice theory and simulation
algorithm. This is followed by a detailed description of
our numerical results. We compare them to the theory with fermions in the fundamental representation. Our results show clear differences between
the two theories -- the symmetric quark theory has a behavior similar
to the fundamental quark theory at strong coupling but deviates substantially
for weak coupling where we observe substantially lighter hadronic masses.

\section{Review of some Theoretical and Phenomenological aspects of Higher Dimensional Representations}
The phase diagram of strongly coupled theories is relevant both theoretically and phenomenologically. By comparing the dynamics of various strongly coupled theories in different regimes we acquire a deeper understanding of non-perturbative dynamics. 
  
\subsection{Conformal Window}  
\begin{figure}[h]
\resizebox{8.5cm}{!}{\includegraphics{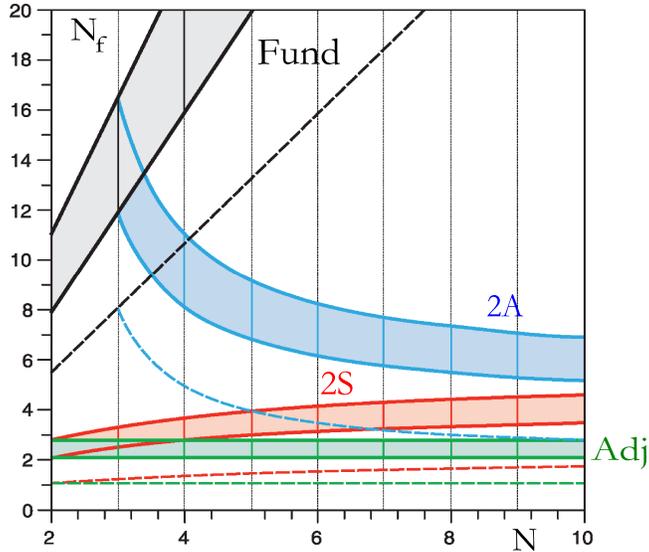}}
\caption{Phase diagram for theories with fermions in the: i) fundamental representation (grey), ii)
two-index antisymmetric (blue), iii) two-index symmetric (red), iv) adjoint
representation (green) as a function of the number of flavors and
the number of colors. The shaded areas depict the corresponding conformal
windows. The upper solid curve represents the loss of
asymptotic freedom, the lower curve loss of chiral 
symmetry breaking. The dashed curves show the existence of a Banks--Zaks fixed point. Picture taken from \cite{Dietrich:2006cm}.}
\label{PH}
\end{figure}

Recently we have completed the analysis of the phase diagram 
of asymptotically free non supersymmetric gauge theories with
at least two Dirac fermions in a given arbitrary representation of the gauge group
as function of the number of
flavors and colors \cite{Dietrich:2006cm}. 
With the exceptions of a few isolated higher dimensional representations below nine colors (fully investigated in \cite{Dietrich:2006cm}) the main phase diagram taken from \cite{Dietrich:2006cm} is sketched in figure \ref{PH}.
The analysis exhausts
the phase diagram for gauge theories with Dirac fermions in arbitrary
representations and it is based on the ladder approximation presented in \cite{Appelquist:1988yc,Cohen:1988sq}.  Further studies of the conformal window and its properties can be found in \cite{Appelquist:1996dq,MY,Gies:2005as}.

In the plot the shaded areas represent the conformal windows for the fundamental representation (grey), two-index antisymmetric (blue), two-index symmetric (red) and adjoint representation (green). {}For each representation the upper solid curve represents the loss of
asymptotic freedom, the lower curve loss of chiral 
symmetry breaking. The dashed curves show the existence of a Banks--Zaks fixed point \cite{Banks:1981nn}. Note how consistently the various
representations merge into each other when, for a specific value of $N$,
they are actually the same representation.

Remarkably the adjoint and the two index-symmetric representation need a very low number of flavors, for an arbitrary number of colors, to be near an infrared fixed point. {}For any number of colors, in the case of the adjoint representation and for two colors in the case of the symmetric representation the estimated critical lower number of flavors above which the theory is already inside the conformal window and hence no chiral symmetry breaking occurs is $N_f^c \sim 2.075$ \cite{Sannino:2004qp,Dietrich:2006cm}. Since this theory walks already for two flavors and hence has also a 
nontrivial chiral dynamics it will be denoted the minimal walking theory. 

The theoretical estimates of the conformal window presented above need to be tested further. The very low number of flavors needed to reach the conformal window makes the minimal walking theories amenable to lattice investigations. 

We also note that the theory with two colors and two Dirac flavors in the fundamental representation is very much below the critical number of flavors needed to develop a non trivial infrared fixed point and hence it features QCD-like dynamics. 

\subsection{Minimal Walking Technicolor}

New strongly interacting
theories can emerge in extensions of the
Standard Model. For example, to avoid
unnaturally large quantum corrections to the mass squared of the elementary Higgs one can replace it by a new strongly coupled fermionic
sector. This is the {\it technicolor} mechanism \cite{TC}. The generation of the masses of the standard model fermions
requires extended technicolor interactions. To avoid large 
flavor-changing neutral currents technicolor theories possessing a sufficient amount of 
walking
\cite{Eichten:1979ah,Holdom:1981rm,Yamawaki:1985zg,Appelquist:an,Lane:1989ej} are needed. 
The simplest of such models which also passes the electroweak
precision tests (like, for example the experimental bounds on the oblique
parameters) has fermions in higher dimensional representations
of the technicolour gauge group
\cite{Sannino:2004qp,Dietrich:2005jn}. We have shown in \cite{Dietrich:2006cm} that the minimal walking theory is also the minimal walking technicolor theory. Such a theory has also a number of desirable features. For example, together with a minimal modification of the SM fermionic matter content it yields a high degree of unification, at the one loop level,  of the SM couplings \cite{Gudnason:2006mk}. Straightforward extensions of the minimal walking theory able to accommodate extended technicolor interactions  can be constructed \cite{Evans:2005pu}.

\section{Lattice Theory}
The lattice action we employ consists of the usual Wilson plaquette
term 
\begin{equation}
S_G=-\frac{\beta}{2}\sum_x\sum_{\mu>\nu}{\rm Re}{\rm Tr}\left(
U_\mu(x)U_\nu(x+\mu)U^\dagger_\mu(x+\nu)U^\dagger_\nu(x)\right) \ ,\end{equation}
together with the Wilson action for two Dirac quarks in the symmetric
representation
\begin{eqnarray}
S_F&=&-\frac{1}{2}
\sum_x\sum_\mu\psib(x)\left(
V_\mu(x)\left(I-\gamma_\mu\right)\psi(x+\mu)+
V^\dagger_\mu(x-\mu)\left(I+\gamma_\mu\right)\psi(x-\mu)
\right)\\
&+&\sum_x \left(m+4\right)\sum_x\psib(x)\psi(x) \ ,
\end{eqnarray}
where the symmetric links are given by
\begin{equation}
V^{ab}_\mu(x)={\rm Tr}\left(S^aU_\mu(x)S^bU^T_\mu(x)\right) \ ,
\end{equation}
and the matrices $S^a,a=1,2,3$ are a basis for the symmetric
representation
\[\begin{array}{ccc}
S^1=\frac{1}{\sqrt{2}}\left(\begin{array}{cc}1&0\\0&1\end{array}\right)\ , \,\,&\,\,
S^2=\frac{1}{\sqrt{2}}\left(\begin{array}{cc}0&1\\1&0\end{array}\right)\ ,\,\,&\,\, 
S^3=\frac{1}{\sqrt{2}}\left(\begin{array}{cc}1&0\\0&-1\end{array}\right) \ .
\end{array}\]
We have simulated this theory over a range of
gauge couplings $\beta=1.5 - 3.0$ and bare quark masses 
$m$ ranging from $ -1.0 < m <1.0 $ on $4^3\times 8$ lattices 
using the usual Hybrid Monte Carlo algorithm
\cite{hmc}. We have focused on determining the critical
line $m_c(\beta)$ needed to approach the continuum limit
and a variety of meson masses and decay constants.
Typically we generate between $400-2000$
$\tau=1$ HMC trajectories. Periodic boundary conditions were used for
all fields.

In tandem with these symmetric quark runs we have also simulated the
theory with fundamental quarks. This allows us to make comparisons at
identical lattice volumes and comparable couplings
and masses and helps highlight the essential differences associated
with the walking dynamics. 

\section{Numerical Results}
\subsection{Pion and rho mass}
We estimate the hadron masses by suitable fits to
corresponding time sliced averaged
correlation functions
\[G_O(t)=\sum_{x,y}<\psib(x,t)\Gamma_O\psi(x,t)\psib(y,0)\Gamma_O\psi(y,0)> \ ,\] 
where
$\Gamma_O=\gamma_5$ for the pion and $\Gamma_O=\gamma_\mu, \mu=1,2,3$ for
the rho (the latter being averaged over spatial directions $\mu$).
In practice we use point sources located at (odd,odd,odd) lattice sites
on the $t=0$ timeslice.
Since our lattices are so small we have simply thrown out the $t=0$ data point
and fitted the remaining correlator to a simple hyperbolic cosine of
the form $a_O\cosh{(m_O(t-L/2))}$ to
estimate the corresponding meson mass $m_O$.

\begin{figure}
\resizebox{7.0cm}{!}{\includegraphics{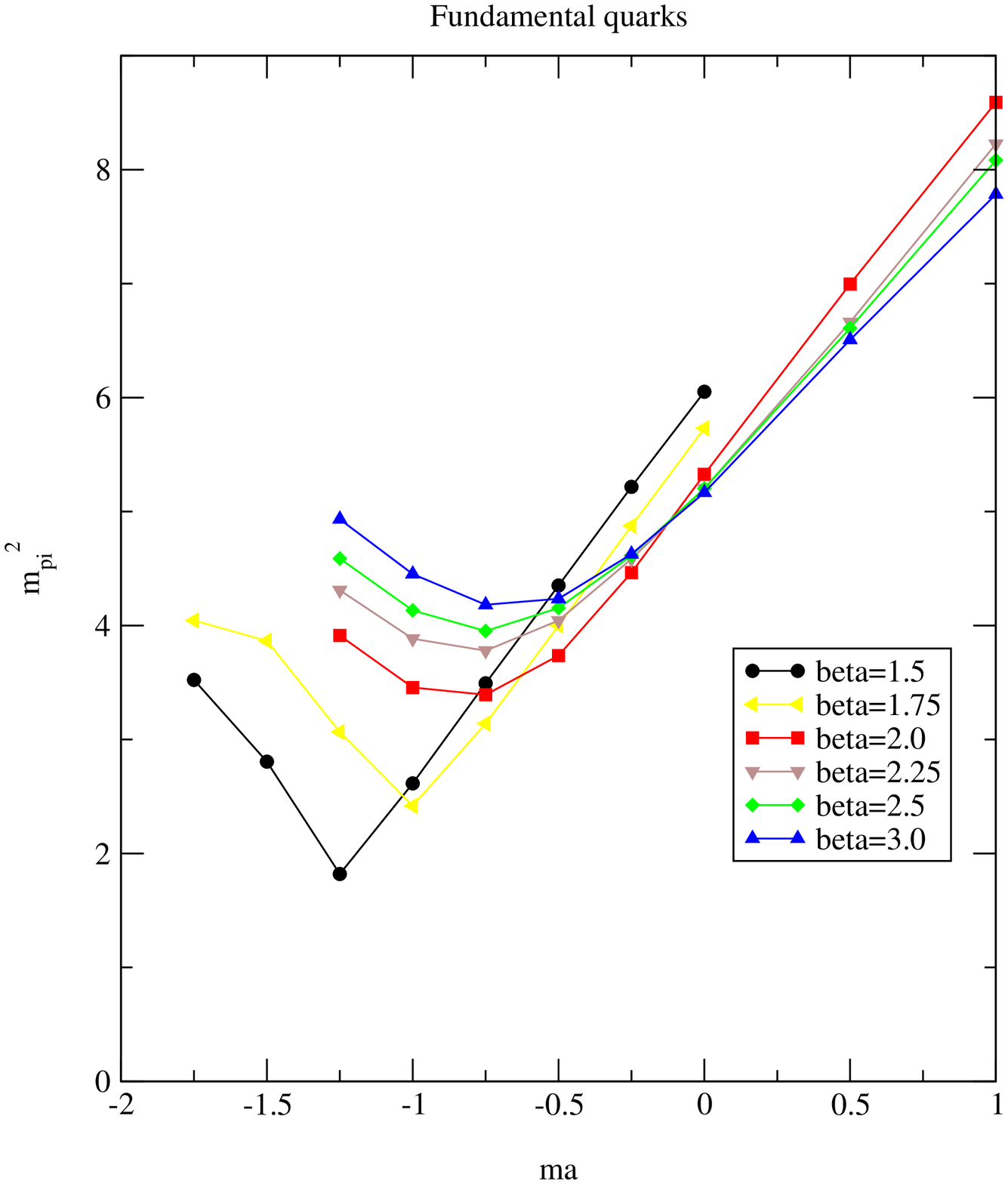}}
\caption{Pion mass squared vs quark mass for fundamental quarks}
\label{pion_fund}
\end{figure}

Consider first the usual case corresponding to taking the quarks to lie
in the fundamental representation of the gauge group.
Figure~\ref{pion_fund} shows curves of the pion mass squared (lattice
units) for
fundamental quarks at several
values of $\beta$ as a function of the bare quark mass $m=Ma$. At each
coupling $\beta$ we observe the usual linear variation of $m_{\pi}^2$ with
bare quark mass $m$. We see that the pion mass attains a minimum value
for some critical quark mass $m_c(\beta)$ which moves towards $m=0$ as
$\beta$ increases. This is similar to the situation in QCD.
Notice that the minimal pion mass {\it increases} with $\beta$ as a result
of finite volume effects.

\begin{figure}
\includegraphics[width=7.0cm,clip=true]{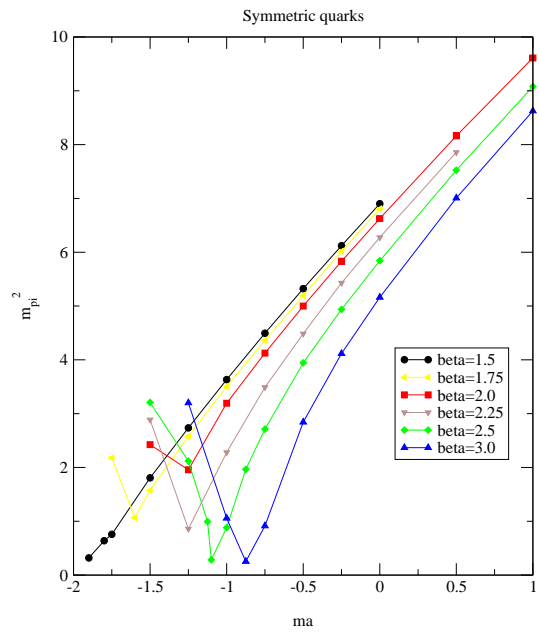}
\caption{Pion mass squared vs quark mass for symmetric quarks}
\label{pion_sym}
\end{figure}

Contrast this behavior with the analogous curves for the symmetric
representation in fig.~\ref{pion_sym} for the same
bare couplings $\beta$ and over the same range of bare quark
mass $m$. Again, a Goldstone
behavior is seen over some range of bare quark mass with
a critical quark mass $m_c(\beta)$
that runs towards the origin as $\beta$ increases.
Notice though that this linear regime appears distorted 
at larger $\beta$ which
we interpret to be the result of strong finite volume effects. This
is consistent with the small values for the pion mass 
$m_\pi\sim 0.5$ observed there. Indeed the minimal
pion mass is a factor of $3-4$ times smaller for the symmetric
pions than pions built from fundamental quarks. The appearance of
this light scale is the first indication that the dynamics
of this theory could be quite different from QCD.

To make this clearer notice that the meson mass in lattice units
is given in general by a pure number
times a non-perturbative lattice scale parameter $\Lambda_{\rm latt}$
which is related to the beta-function in the usual way
\[\Lambda_{\rm latt}=\Lambda_{\rm phys}a\sim e^{-\int \frac{dg}{\beta(g)}} \ .
\] 
This scale is small if the theory lies close to
a zero of the beta-function.
Furthermore, finite size effects are governed by
the quantity $\Lambda_{\rm latt}L$. For fixed $L$ the theory with
the smaller $\Lambda_{\rm latt}$ might be expected to
exhibit larger finite size
effects -- as we observe for the symmetric quark theory. 

\begin{figure}
\centering
\begin{tabular}{cc}
\resizebox{6.0cm}{!}{\includegraphics{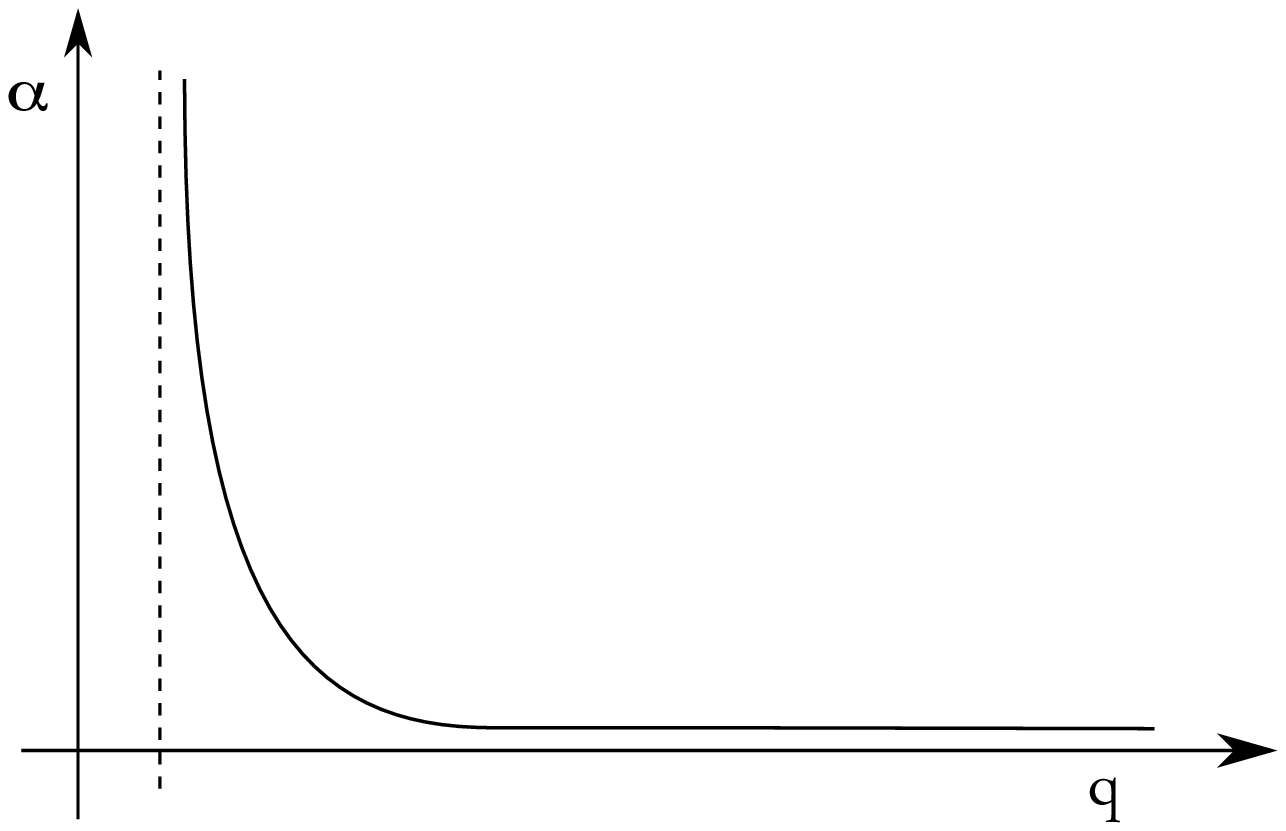}} ~~~&~~ \resizebox{6.0cm}{!}{\includegraphics{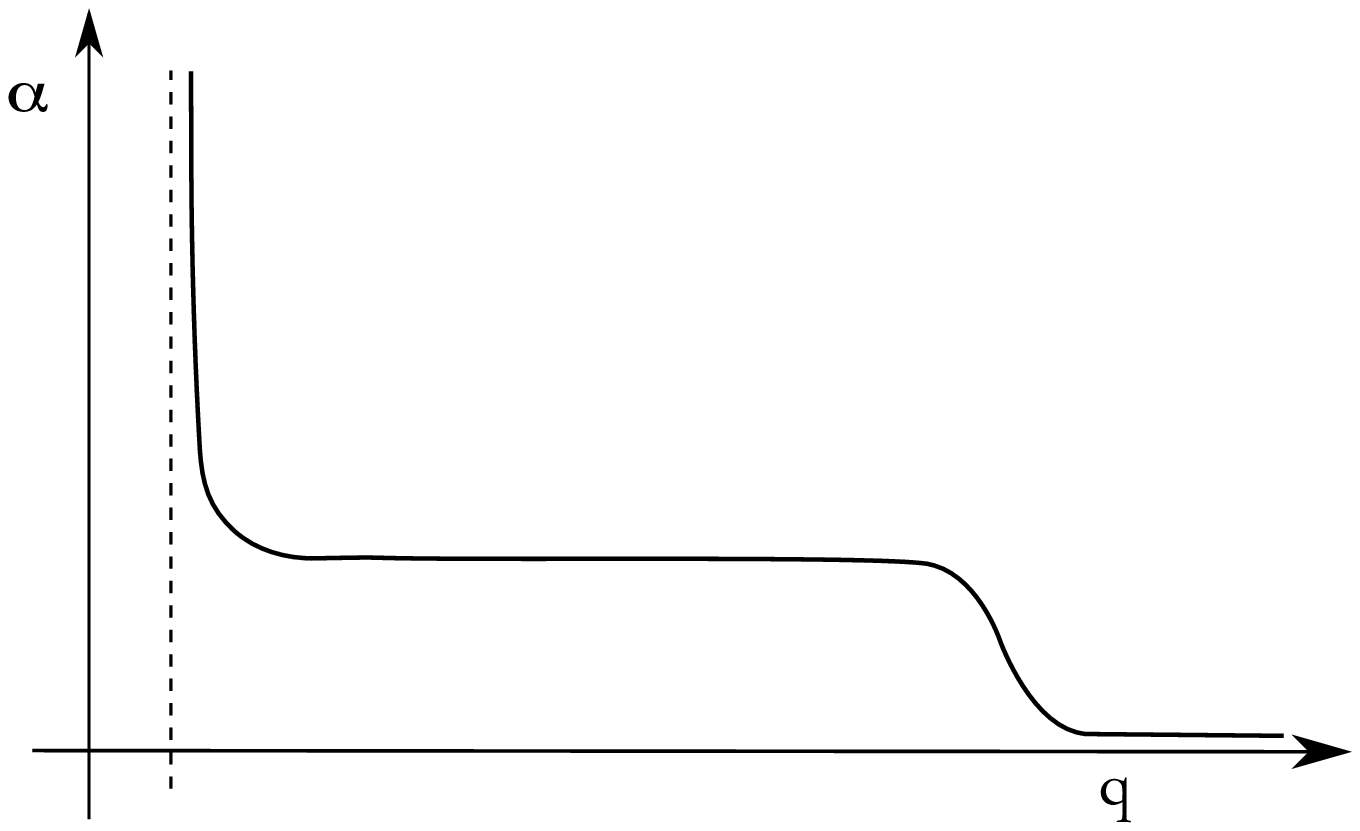}} \\&\\
&~~~~\resizebox{6.0cm}{!}{\includegraphics{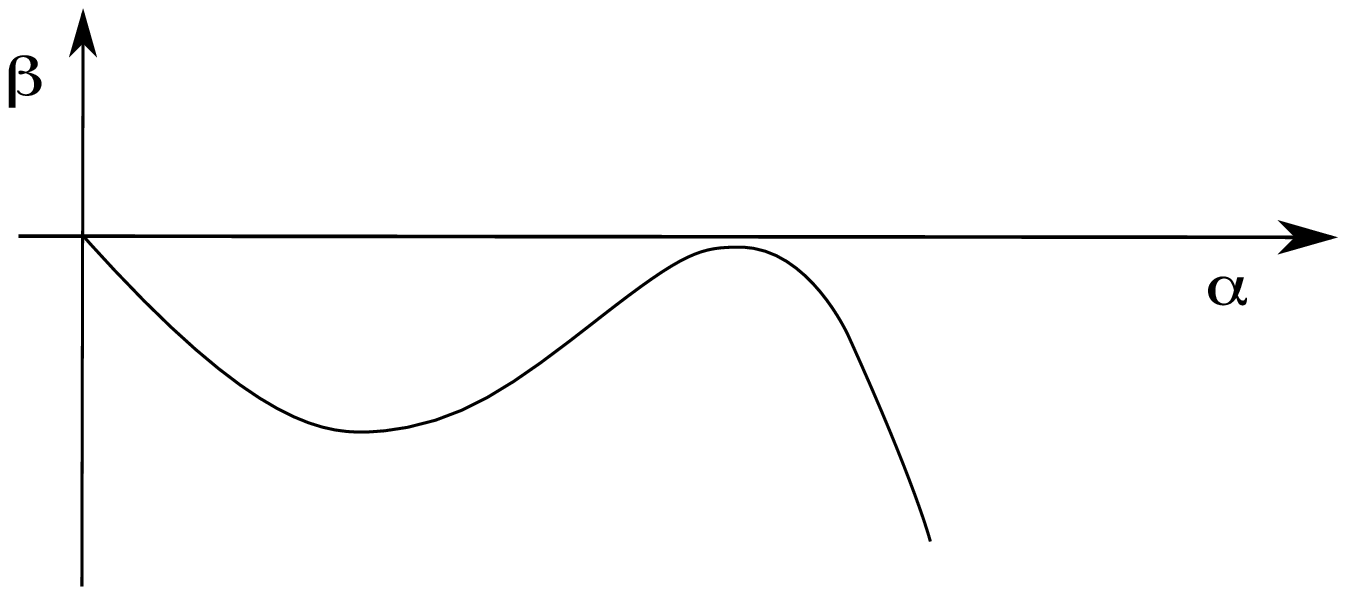}} 
\end{tabular}
\caption{Top Left Panel: QCD-like behavior of the coupling constant as function of the momentum (Running). Top Right Panel: Walking-like behavior of the coupling constant as function of the momentum (Walking). Bottom Right Panel: Cartoon of the beta function associated to a generic walking theory.}
\label{walkbeta}
\end{figure}

However, perhaps the most striking
difference between these symmetric plots and their fundamental cousins
is the {\it non-monomtonic} behavior of the pion mass with increasing
$\beta$. For small $\beta$ the minimal pion mass increases with $\beta$
in a way which is similar to the fundamentals. However for $\beta >2$
the pion mass {\it decreases} with further increases in $\beta$. This
is consistent with the existence of a beta-function
(at zero quark mass) which resembles that
of QCD at strong coupling but is small for couplings less than some
characteristic value. Such a beta-function corresponds to the case of
walking dynamics.  Fig~\ref{walkbeta} gives a schematic
picture of the beta-function and related coupling constant evolution
for such a walking theory. 

\begin{figure}
\resizebox{7.0cm}{!}{\includegraphics{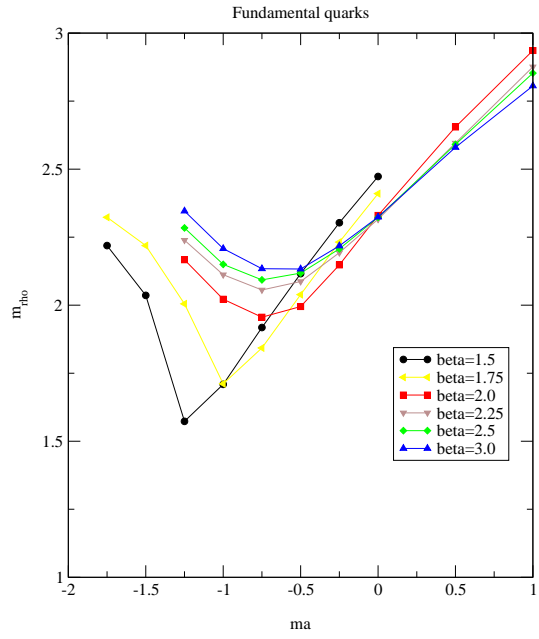}}
\caption{Rho mass vs quark mass for fundamental quarks}
\label{rho_fund}
\end{figure}

\begin{figure}
\resizebox{7.0cm}{!}{\includegraphics{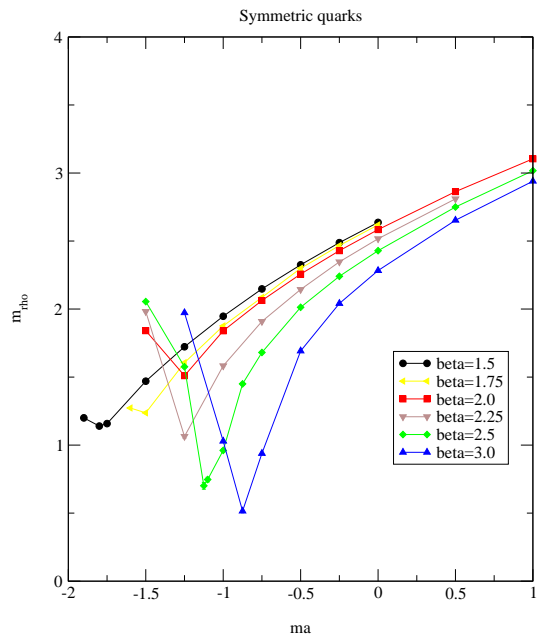}}
\caption{Rho mass vs quark mass for symmetric quarks}
\label{rho_sym}
\end{figure}

These differences are also seen in the rho meson. Figure~\ref{rho_fund}
and fig.~\ref{rho_sym} 
show the rho mass (in lattice units) as a function of quark mass
over the same range of couplings $\beta$ for
both fundamental and symmetric quarks. Again a minimal rho mass
is seen which coincides approximately with the minimal pion mass
and moves towards the origin as $\beta$ increases.
In our later analysis we have used the
minimal rho mass rather than that of the pion when trying to get an estimate of the critical line.
In the case of fundamental quarks the observed rho mass increases 
montonically with $\beta$
but again for symmetrics the minimal rho mass shows a non-monotonic
behavior with $m_\rho$ initially increasing and then falling as
$\beta$ increases. The similar behavior of the rho mass to the pion mass
for symmetric quarks lends strong support 
to the walking hypothesis since the near
conformal dynamics will suppress the mass of all hadron states - not
just the pi meson.

\subsection{Pion decay constant} 
\begin{figure}
\resizebox{7.0cm}{!}{\includegraphics{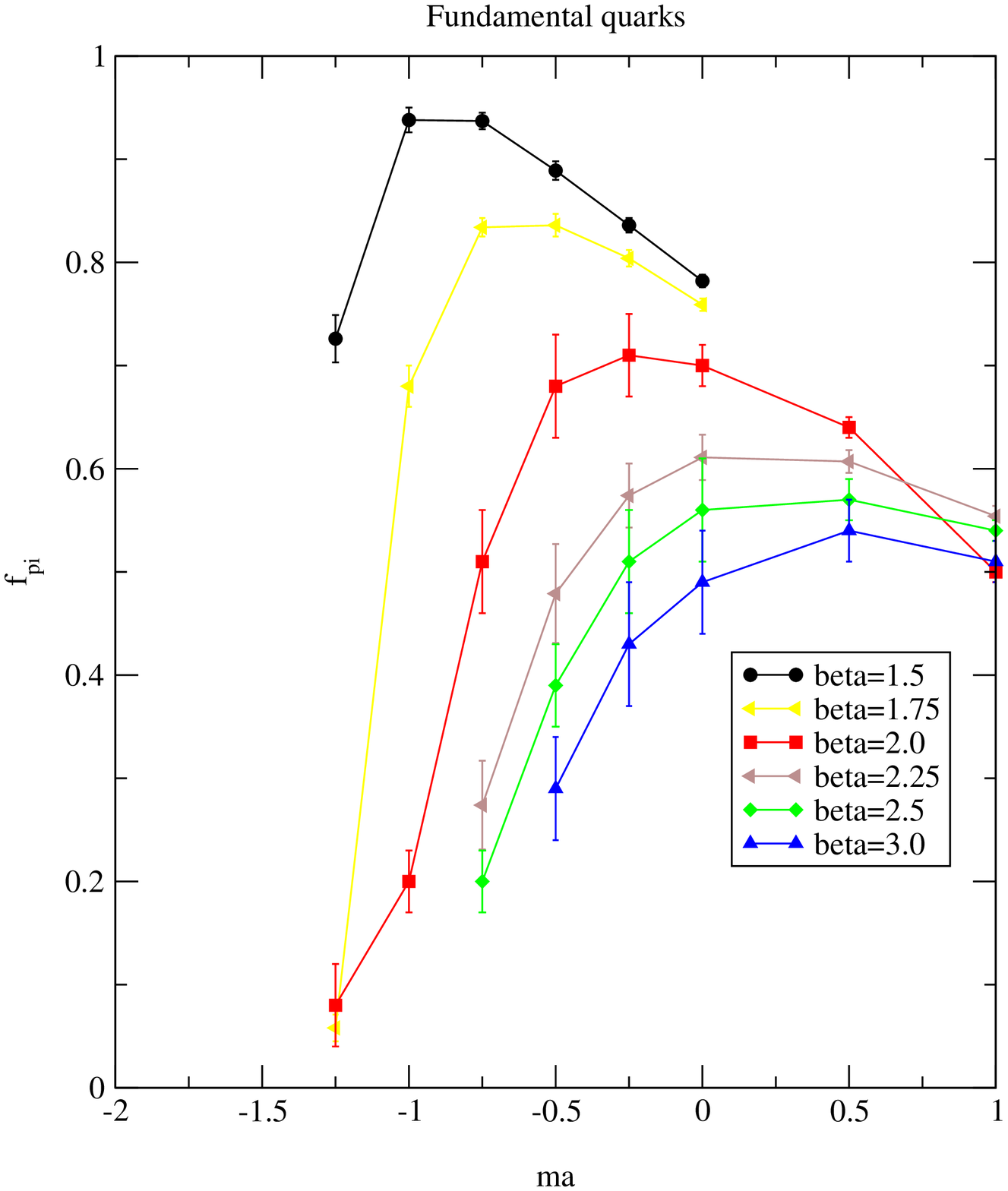}}
\caption{$f_\pi$ vs quark mass for fundamental quarks}
\label{fpi_fund}
\end{figure}

\begin{figure}
\resizebox{7.0cm}{!}{\includegraphics{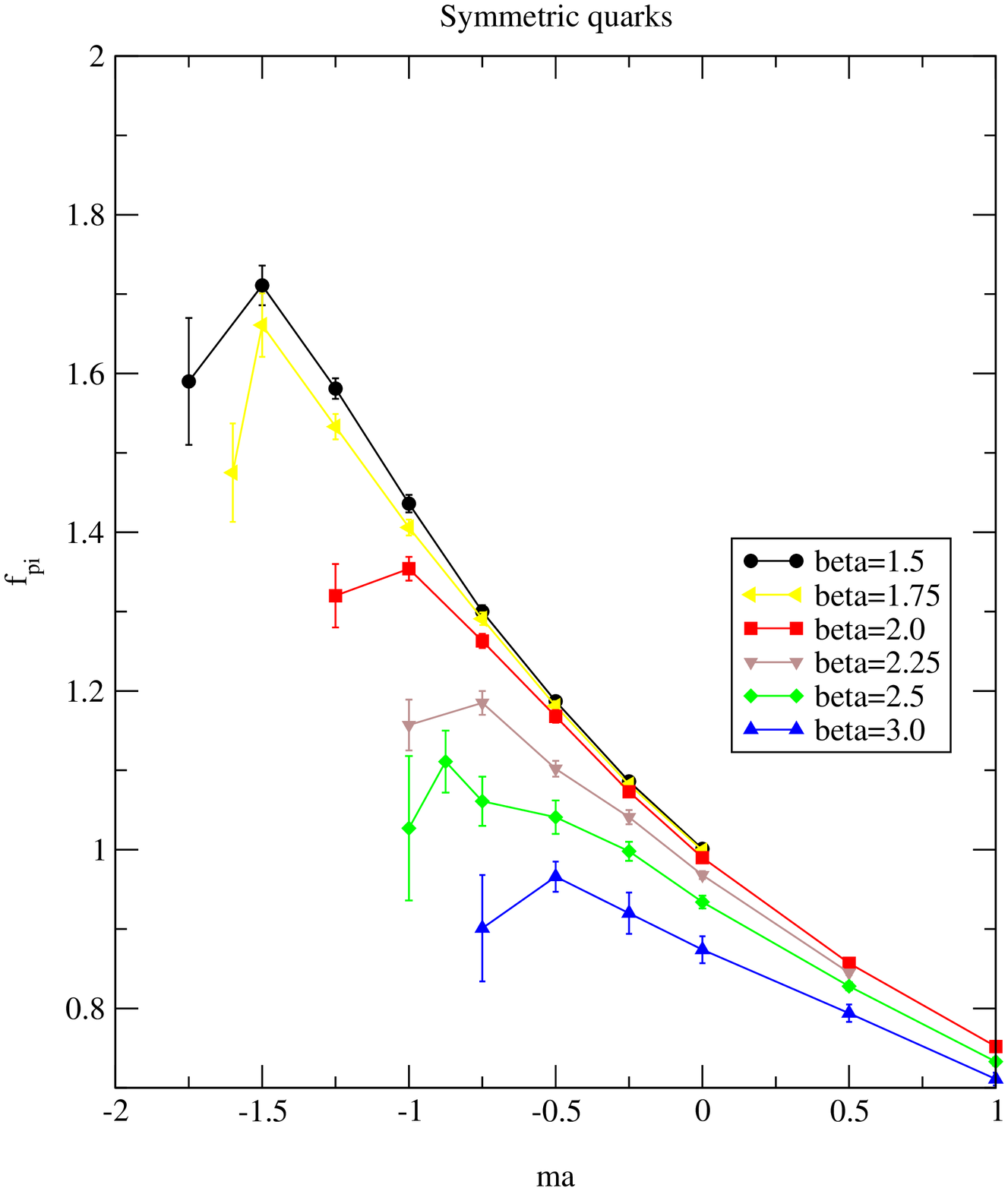}}
\caption{$f_\pi$ vs quark mass for symmetric quarks}
\label{fpi_sym}
\end{figure}

To compute the pion decay constant we have also measured the
axial correlator defined by
\begin{equation}
G_A(t)=\sum_{x,y}<\psib(x,t)\Gamma_0\Gamma_5\psi(x,t)
                  \psib(y,0)\Gamma_5\psi(y,0)> \ .
\end{equation}
As for the hadron correlators we discard the $t=0$ data point
and fit the remaining
function to a hyperbolic sine function $a_A\sinh{(m_\pi(t-L/2))}$. We observe
that the pion mass extracted from this fit is indeed consistent with that
derived from the pion correlator in the physical regime $m>m_c(\beta)$. 
An expression for pion decay constant (in lattice units) is then given by
\cite{decay}
\begin{equation}
\frac{f_\pi}{Z}=\frac{a_A}{\sqrt{a_\pi m_\pi}}e^{m_\pi L/4} \ .
\end{equation}
We do not attempt in this paper to determine the renormalization
constant $Z$ and hence all results for $f_\pi$ are only derived up
this factor. 
Figure~\ref{fpi_fund} shows this quantity for a range of $\beta$ and quark
masses $m\ge m_c(\beta)$ for the theory with quarks in the fundamental.
We see that $f_\pi$ increases as the critical line is approached with
$f_\pi$ taking on a maximal value there before falling rapidly to
zero in
the phase with $m<m_c(\beta)$. The value of $f_\pi$ along the
critical line falls as $\beta$ is increased and the lattice spacing
reduced. 
Equivalent data for symmetric quarks
is shown in fig.~\ref{fpi_sym}. Notice that the maximal
value of $f_\pi$ is about twice that found for fundamentals 
(where both are measured in lattice units) but otherwise
the picture is qualitatively similar. Presumably the fact that
$f_\pi$ is {\it larger} rather than smaller for symmetric quarks as compared
to fundamentals, 
contrary to naive expectations, could be
related to the differing renormalization
constants $Z$ in the two cases.

\subsection{Scaling}

\begin{figure}
\resizebox{7.0cm}{!}{\includegraphics{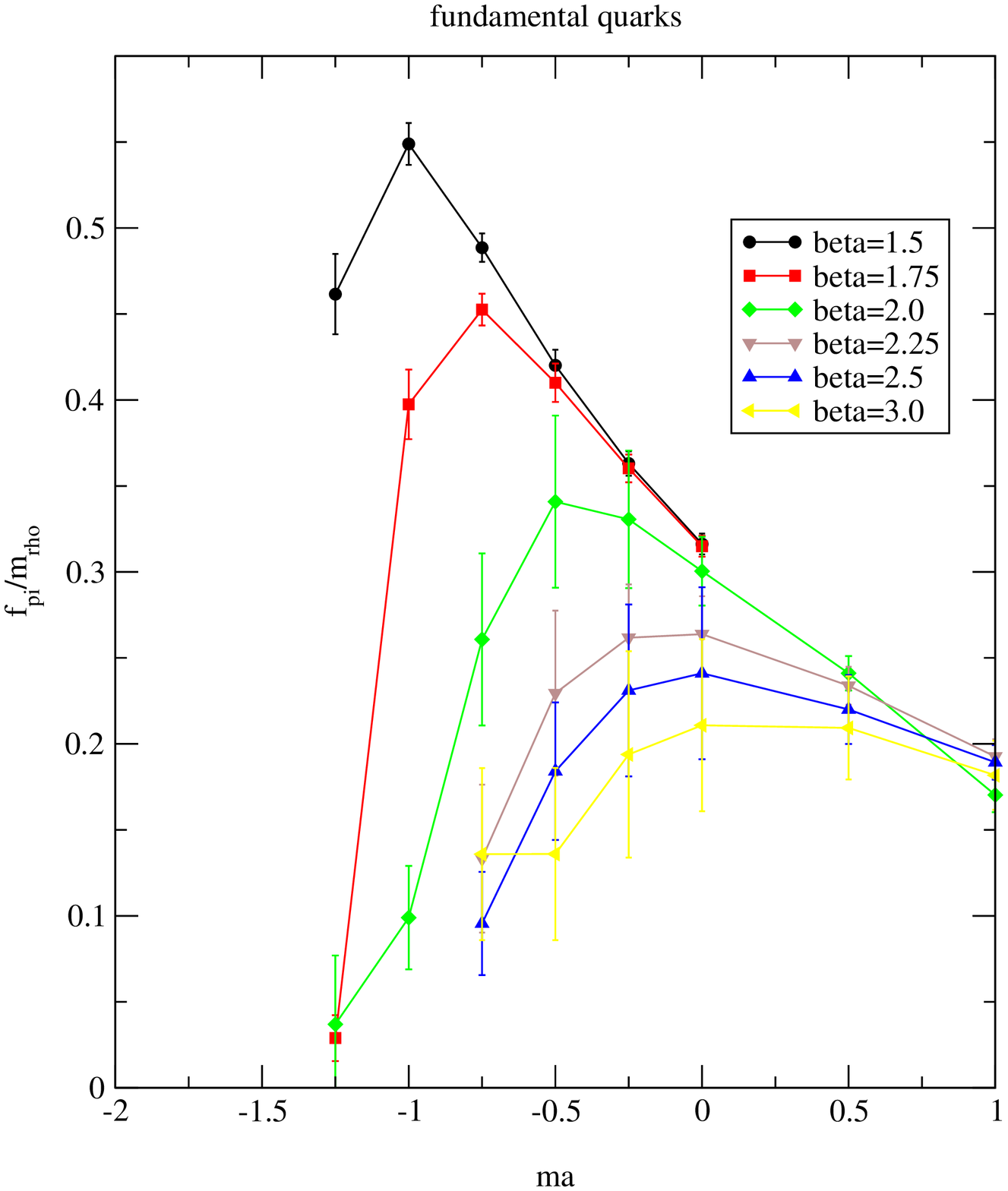}}
\caption{$\frac{f_\pi}{m_\rho}$ vs quark mass for fundamental quarks}
\label{fpi_rho_ratio_fund}
\end{figure}

\begin{figure}
\resizebox{7.0cm}{!}{\includegraphics{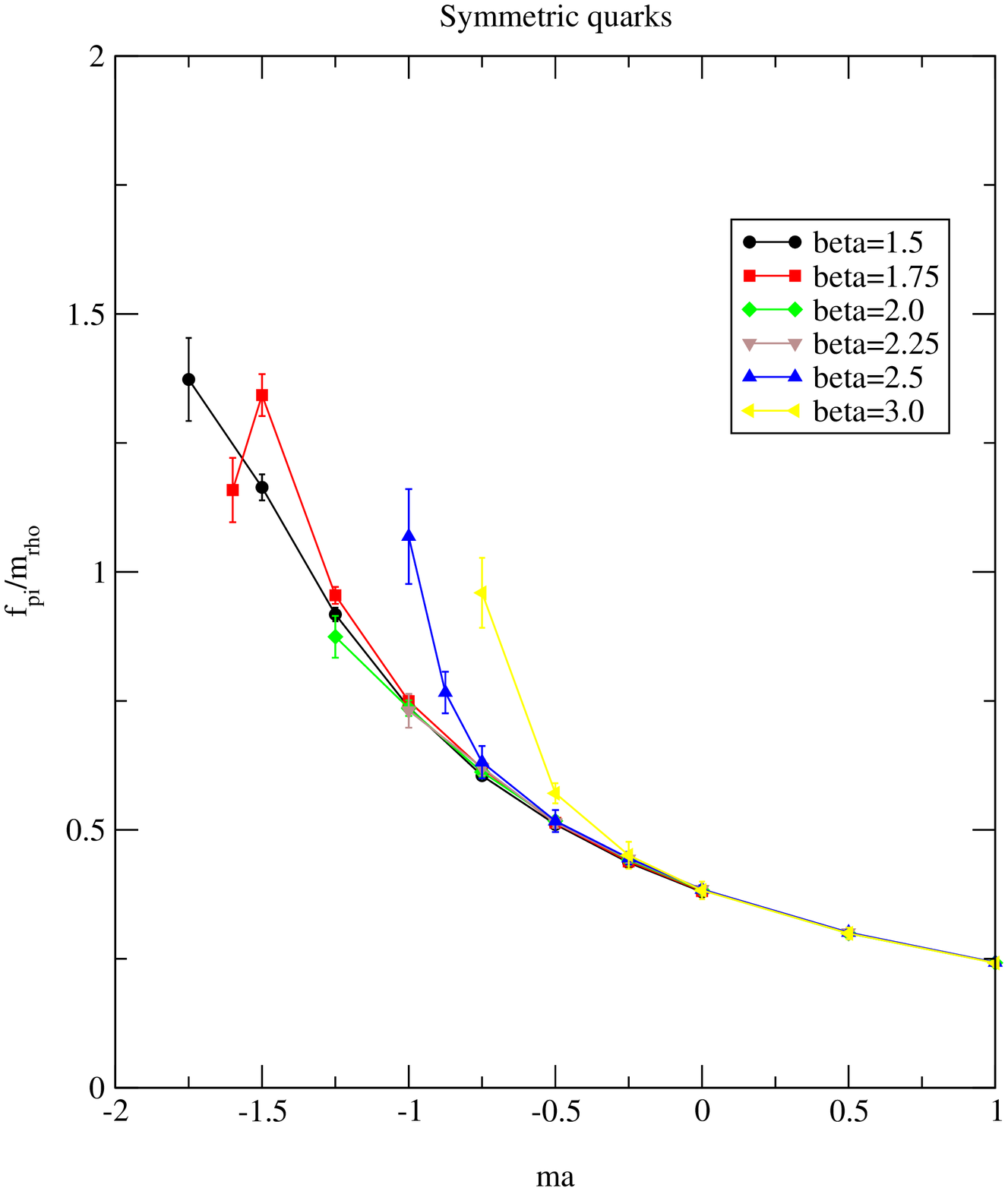}}
\caption{$\frac{f_\pi}{m_\rho}$ vs quark mass for symmetric quarks}
\label{fpi_rho_ratio_sym}
\end{figure}

To examine the continuum limit in greater detail it is useful to
examine dimensionless quantities as we scan in $\beta$ and $m$. Thus
we have examined the ratio $\frac{f_\pi}{m_\rho}$ which is plotted
in figs.~\ref{fpi_rho_ratio_fund} and \ref{fpi_rho_ratio_sym} over the
same range of couplings and masses. For both types of representations
we see that $\frac{f_\pi}{m_\rho}$ increases as the quark mass is tuned
towards the critical line. Furthermore there is evidence that for large
enough quark mass the curves for different bare coupling $\beta$ lie
on top of one another - this was not true for the bare $f_\pi$. 
Notice also that the values of $\frac{f_{\pi}}{m_\rho}$ for the
two representations are much closer in this region
of parameter space.

To extract continuum physics we should look for scaling -- dimensionless
quantities should become independent of coupling along the
critical line as $\beta\to\infty$. We see some evidence for
this away from the critical line where the sets of curves for
different bare coupling $\beta$ lie on top of one another. This
approximate collapse of the data onto a single scaling curve at
intermediate quark masses
happens for both representations of dynamical quark.

However, if we 
look at the values of $\frac{f_{\pi}}{m_\rho}$ for 
fundamental quarks near the critical quark mass
$m\sim m_c$  we see that they
{\it decrease} with increasing $\beta$. This is to be attributed to
finite volume effects -- as the lattice spacing decreases the physical
volume also decreases since we use a fixed lattice volume. The rising
rho mass with physical volume then leads to a falling value for
this ratio.
Notice also that the $\frac{f_\pi}{m_\rho}$ falls {\it below} the
scaling curve as $m\to m_c$. This should be contrasted
with the case of the symmetric quarks where the situation is reversed -- the
value of the ratio $\frac{f_\pi}{m_\rho}$ increases above the 
scaling curve as the quark mass is tuned towards its critical value. 

\begin{figure}
\resizebox{7.0cm}{!}{\includegraphics{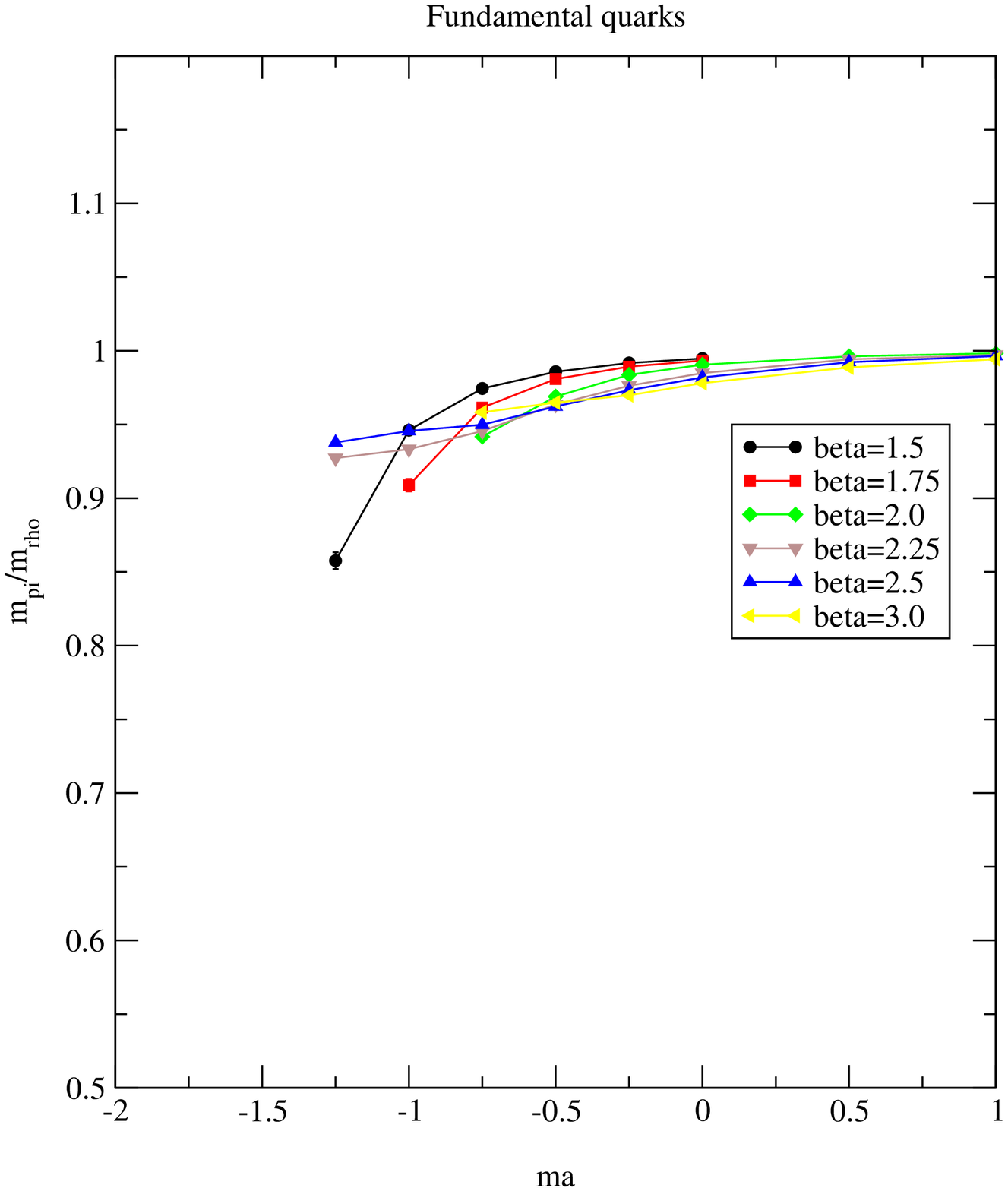}}
\caption{$\frac{m_\pi}{m_\rho}$ vs quark mass for fundamental quarks}
\label{pi_rho_ratio_fund}
\end{figure}

\begin{figure}
\resizebox{7.0cm}{!}{\includegraphics{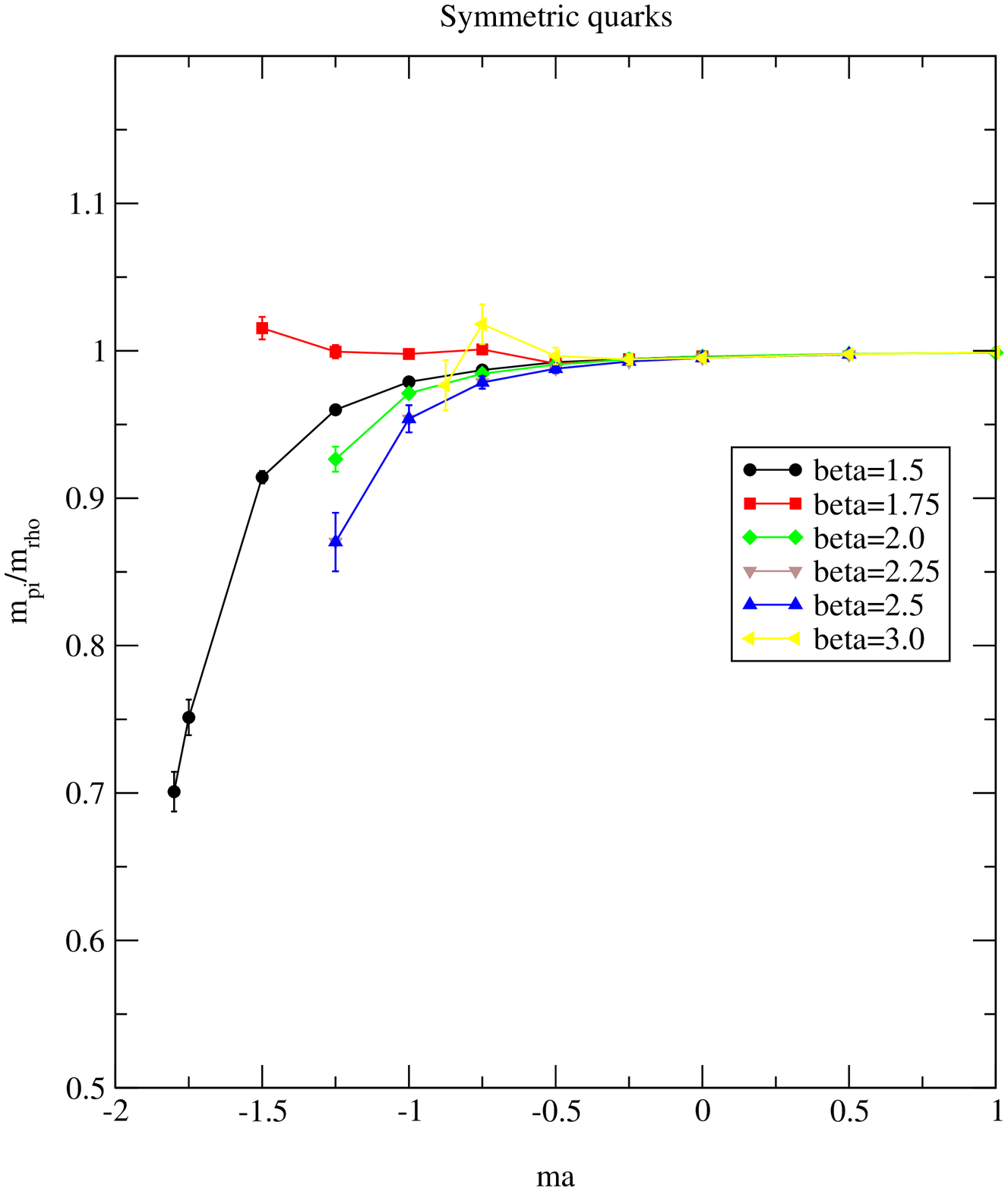}}
\caption{$\frac{m_{\pi}}{m_\rho}$ vs quark mass for symmetric quarks}
\label{pi_rho_ratio_sym}
\end{figure}

Figs.~\ref{pi_rho_ratio_fund} and \ref{pi_rho_ratio_sym} show plots of
the dimensionless ratio of pion to rho mass for the two
theories. Both show values which decrease from unity as the critical
line is approached. There is some evidence that the theory with
symmetric quarks exhibits a value of the ratio closer to unity as
the critical line is approached and that this limiting value
is rather insensitive to the bare coupling for large $\beta$ -- a hint perhaps
of walking dynamics.

\section{Summary}
In this paper we report on the results of lattice simulations of 
$2$ color QCD coupled to $2$ flavors of quarks in the symmetric representation
of the gauge group. These studies are motivated by the idea that this
theory lies close to a strongly coupled conformal field theory 
associated to a new I.R. attractive zero of the beta function.
The resultant theory is expected to exhibit non-QCD like dynamics 
and a slow evolution of the coupling constant over some range of
scales.

To search for these effects we have simulated both the symmetric
quark model and  
its counterpart employing fundamental quarks on identical
lattice volumes and for comparable couplings
and quark masses.
In our simulations we observe several features which distinguish the
symmetric from the fundamental representation. The most obvious of these
can be seen in the hadron masses. While the symmetric pion and rho behave
in a qualitatively similar way to their fundamental counterparts at
strong coupling they depart from this behavior for $\beta  \gtrsim 2.0$.
In this regime their (lattice) masses decrease with further increases in
the bare
coupling $\beta$ attaining values substantially smaller than those
seen in the theory with fundamental quarks. This observation is
consistent with the existence of a light scale $\Lambda_{\rm latt}$
attributable to walking dynamics.

However, our results can also be interpreted as 
evidence that for sufficiently light
symmetric quarks and large $\beta$
the 2 color theory has {\it already} entered a conformal
phase in which the quarks are deconfined at zero temperature. This could
explain the lack of linear scaling of $m^2_\pi$ with $m$ at large
$\beta$ since in such a phase chiral symmetry would
be restored. If so, it is unclear whether this is the result of strong
finite volume effects and will disappear on larger lattices or
is a genuine bulk transition occuring for
some $\beta_c$. If the latter scenario turns out to be true it would
imply that the critical number of flavors needed to access 
the conformal phase is smaller than the perturbative estimate
$N_f^c=2.075$. Further work on larger
lattices will be needed to answer
this question definitively.

In the case of
fundamental quarks a much larger number of flavors is needed
to approach the conformal window. Nevertheless, a small suppression
of chiral symmetry breaking effects 
was observed in \cite{mawhinney}. More recently a study of QCD
with fundamental quarks showed evidence for an intermediate
conformal phase for $7<N_f<17$ \cite{iwasaki}. 

Exploratory simulations of this model with staggered quarks at finite
temperature were conducted earlier \cite{kogut}. Our results at zero
temperature complement and extend that work. 

Clearly our lattices are very small and so our results should be seen 
primarily as providing motivation for a study on larger lattices. 
Going to larger lattices will allow us to understand better the
finite volume effects clearly visible in this work, will allow
for a more careful study of the hadron spectrum and should
allow simulations to be undertaken at smaller cut-off. With a larger
lattice it should also be possible to see direct evidence for
walking by measuring the running of a renormalized coupling 
extracted from either the static quark potential or
using Schr\"{o}dinger functional techniques. This work
is currently underway. 

 \acknowledgments
We are happy to thank L. Del Debbio, M. T. Frandsen and L. Giusti for helpful discussions. The work of S.C. is support in part by DOE grant DE-FG02-85ER40237 while F.S. is supported by the Marie Curie Excellence Grant under contract MEXT-CT-2004-013510 as well as the Danish Research Agency. 

The numerical work was carried out using USQCD resources at Fermilab.

\vfill
\newpage

\vspace{1cm}

\vspace{1cm}

\vspace{1cm}

\vspace{1cm}

\vspace{1cm}

\vspace{1cm}

\vspace{1cm}

\vspace{1cm}

\vspace{1cm}

\vspace{1cm}

\end{document}